\documentclass[a4paper,11pt]{article}
\usepackage{latexsym}
\usepackage{epsfig}
\usepackage[figuresfirst,nomarkers]{endfloat}
\usepackage{caption2}

\def\bg#1{\mbox{\boldmath$#1$}}

\DeclareGraphicsExtensions{.eps}

\newcommand{\beq}{\begin{eqnarray}}
\newcommand{\eeq}{\end{eqnarray}}
\newcommand{\be}{\begin{eqnarray*}}
\newcommand{\ee}{\end{eqnarray*}}

\newcommand{\D}{{\cal D}}
\newcommand{\Pom}{{\hspace{ -0.1em}I\hspace{-0.25em}P}}
\newcommand{\vphot}{{\gamma^*}}
\def\lsim{\raise0.3ex\hbox{$<$\kern-0.75em\raise-1.1ex\hbox{$\sim$}}}
\def\gsim{\raise0.3ex\hbox{$>$\kern-0.75em\raise-1.1ex\hbox{$\sim$}}}

\title{{\bf Gluon shadowing in the Glauber-Gribov model at HERA}}
\author{K.~Tywoniuk\footnote{{\it Corresponding author}:
    \texttt{konrad@fys.uio.no}} $ ^1$, I.~C.~Arsene$^1$,
  L.~Bravina$^{1,2}$, \\ 
  A.~B.~Kaidalov$^3$ and E.~Zabrodin$^{1,2}$ \\ \\ 
  $^1$ Department of Physics, University of Oslo\\
  0316 Oslo, Norway \\
  $^2$ Institute of Nuclear Physics, Moscow State University\\
  RU-119899 Moscow, Russia\\
  $^3$ Institute of Theoretical and Experimental Physics\\
  RU-117259 Moscow, Russia}

\date{}

\begin{document}
  
\vspace{0.5cm}
\maketitle

\abstract{We calculate shadowing using new data on the gluon density
  of the Pomeron recently measured with high precision at HERA. The
  calculations are made in a Glauber-Gribov framework and Pomeron
  tree-diagrams are summed up within a unitarity-conserving procedure.
  The 
  total cross section of $\vphot A$ interaction is then found in a
  parameter-free description, employing gluon diffractive and
  inclusive distribution functions as input. A strong shadowing effect
  is obtained, in good agreement with several other models. Impact
  parameter dependence of gluon shadowing is also presented.}
\par
\vspace{0.3cm}
\begin{centering}
\end{centering}
\par
\vspace{0.3cm}
\begin{centering}
\end{centering}

\newpage

\section{Introduction}\label{sec:intro}

Nuclear shadowing is a well-established phenomenon, which attracts
attention of both experimentalists and theoreticians. It was found
(see \cite{Arn94,GST95} and references therein) that the inclusive
nuclear structure function is smaller in nuclei than in a free nucleon
at small values of the Bjorken variable $x \leq 0.01$. The nature of
shadowing, emerging e.g. in deep inelastic scattering (DIS),
can be well understood in terms of multiparticle scattering in the
target rest frame. The incoming photon is represented as a
superposition of gluons, quarks, anti-quarks and their bound states.
At high bombarding energies the photon converts into $q\bar{q}$-pair
long before the target, and its hadronic component interacts coherently
with several nucleons of the target nucleus. This process leads to
absorption and, therefore, to nucleon shadowing (for a review see
e.g. \cite{Arm06}). Among the
important consequences of the phenomenon is, for instance, severe
reduction of particle multiplicity in heavy-ion collisions at LHC
energies ($\sqrt{s} = 5.5$\, TeV), since multiple scattering is
connected to diffraction \cite{Cap98,Fr02,FSW05}.
The effect can be further decomposed onto the quark shadowing and the
gluon shadowing; the latter provides largest uncertainties in the
theory and is a subject of our present study.

In recent years a lot of interest has been generated about the
possibility of parton saturation in the nuclear wave function at the
smallest $x$ accessible at HERA, and there is an ongoing discussion if
the same effects can be observed for hadron-nucleus and
nucleus-nucleus collisions at
RHIC. In this paper, we will focus mostly on the low-$x$ effects
mentioned above. Understanding the so-called cold nuclear effects in
hadron-nucleus collisions serves also as a
baseline for the correct treatment of possible final state effects in
nucleus-nucleus collisions, e.g. high-$p_T$ particle
suppression and heavy-flavor production at RHIC.

Our starting point is noticing that a significant change in the
underlying dynamics of a hadron-nucleus
collision takes place with growing energy of the incoming particles.
At low energies, the total cross section is well described within
the probabilistic Glauber model \cite{glauber}, which only takes
into account elastic rescatterings of the incident hadron on the
various nucleons of the target nucleus. Elastic scattering is described by
Pomeron exchange. At higher energies, $E > E_{crit} \sim
m_{\scriptstyle{N}} \mu R_A$ ($\mu$ is a characteristic hadronic scale,
$\mu \sim 1$GeV, and $R_A$ is the radius of the nucleus) corresponding to
a coherence length
\beq 
\label{eq:cohlenght} 
l_C \;=\; \frac{1}{2 \, m_N \, x} \;, 
\eeq
the typical hadronic fluctuation length can become of the order of,
or even bigger than, the nuclear radius and there will be
coherent interaction of constituents of the hadron with several
nucleons of the nucleus. The sum of all diagrams was calculated by
Gribov \cite{gribov1,gribov2}, which corrected the Glauber series by
taking into account the diffractive
intermediate states in the sum over 
subsequent rescatterings. The space-time picture analogy to the
Glauber series is nevertheless lost, as the interactions with
different nucleons of the nucleus occurs nearly simultaneous in time.
The phenomenon of coherent multiple scattering is referred to as
shadowing corrections.

An additional effect which comes into play at high energies, is the
possibility of interactions between soft partons of the different
nucleons in the nucleus. In the Glauber-Gribov model this
corresponds to interactions between Pomerons. These diagrams are called
enhanced diagrams \cite{kaidalov_qm}, and can also be understood as
interactions between strings formed in the collision. 
Actually, the necessity to include such diagrams at high
energies can be related to unitarization of the total cross section.
There is a connection between these effects and saturation
effects already mentioned earlier.

The Glauber-Gribov model is described and a unitarity-conserving
procedure for finding the total cross
section of $\vphot$-nucleus ($\vphot A$) interaction,
which corresponds to summing up Pomeron fan-diagrams, is presented in
Sec.~\ref{sec:model}. Further we will concentrate on new and
interesting 
data on gluon diffractive distribution function, which we will
describe in Sec.~\ref{sec:param}. The results for gluon shadowing
are presented in Sec.~\ref{sec:results}, and our conclusions are drawn
in Sec.~\ref{sec:concl}. 

\section{The Model}\label{sec:model}  
\begin{figure}[hb]
  \begin{center}
    \includegraphics[scale=1.]{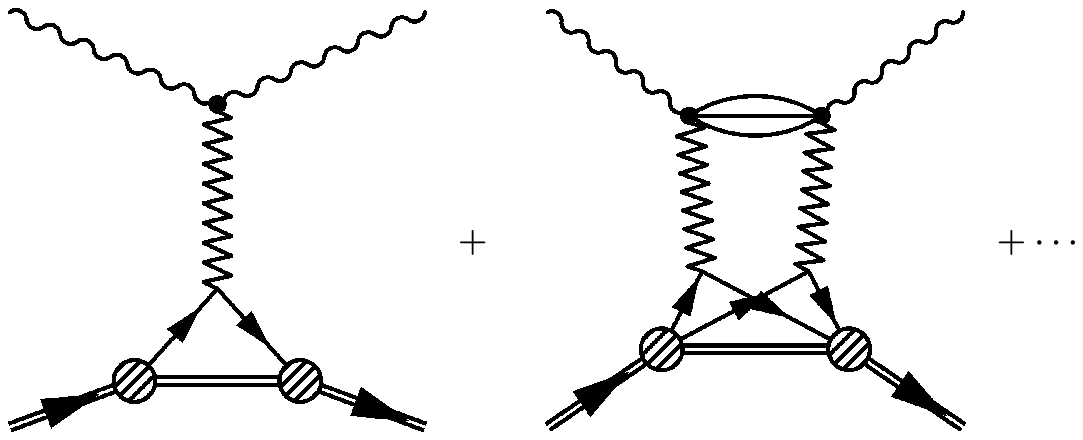}
    \caption{The single and double scattering
      contribution to the total $\vphot N$ cross section.}
    \label{fig:diag1}
  \end{center}
\end{figure}

We consider the nucleus as a set of nucleons, in the spirit of the
Glauber model. The elastic $\vphot A$ scattering amplitude
can then be written as the sum of diagrams shown in
Fig.~\ref{fig:diag1}, i.e. as multiple $\vphot$-nucleon ($\vphot N$) scattering
diagrams with Pomeron exchange \cite{Cap98,armesto03}. The 
contribution from $1, 2...$ scatterings
\beq
\sigma_{\vphot A} \;=\; A\sigma_{\vphot N} \,+\, \sigma_{\vphot A}^{(2)} \,+\, ...\;,
\label{eq:sum}
\eeq
should be summed up to obtain the total cross section. In
Eq.~(\ref{eq:sum}), the first term simply equals to the Glauber 
elastic contribution and subsequent terms describe multiple interactions of the
incoming probe with the 
nucleons in the target nucleus.

\begin{figure}[ht]
  \begin{center}
    \includegraphics[scale=1.]{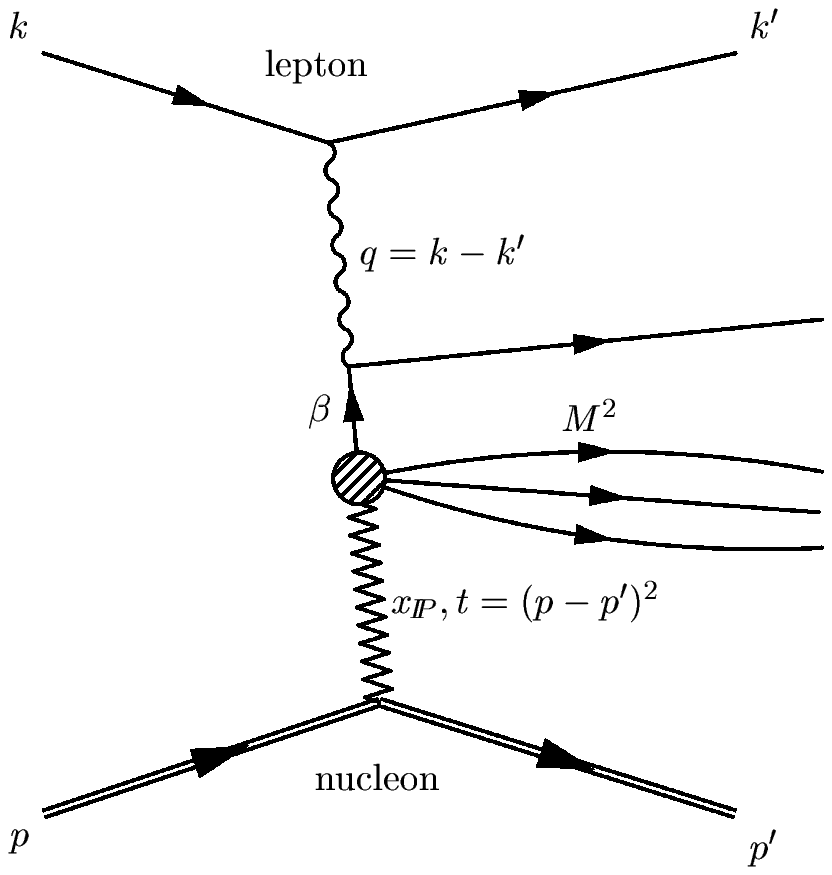}
    \caption{DDIS kinematical variables in the
      infinite momentum frame.}
    \label{fig:diag2} 
  \end{center}
\end{figure}

The multiparticle content of subsequent diagrams in
Fig.~\ref{fig:diag1} is given by
AGK cutting rules \cite{agk}, where the intermediate states are
on-shell. The cut contribution of the double rescattering diagram can be
expressed in terms of diffractive
deep inelastic scattering (DDIS). The usual
variables for DDIS: $Q^2, x, M^2$ and $t$, or $x_\Pom$, are
shown in Fig.~\ref{fig:diag2}. The variable $\beta =
\frac{Q^2}{Q^2+M^2} = x /x_\Pom$ plays the same role for the Pomeron
as the Bjorken variable, $x$, for the nucleon. 
We assume that the 
amplitude of the process is purely imaginary; this is justified for a
value of the Pomeron intercept close to unity \cite{cap01}. The contribution from
the second term in Eq.~(\ref{eq:sum}) to
the total $\vphot A$ cross section is given by \cite{Cap98}
\beq
\label{eq:sec}
\sigma^{(2)}_{\vphot A} \;=\;& & \; -4\pi A(A-1) \times \nonumber \\
& &\times \;\int \mbox{d}^2b \, T_A^2 (b)
\int_{M^2_{min}}^{M^2_{max}} \mbox{d}M^2 \, \left[ \frac{\mbox{d}
    \sigma^{\D}_{\vphot \scriptscriptstyle{N}} (Q^2, x_\Pom, \beta)}{\mbox{d} M^2 \,
    \mbox{d}t}\right]_{t=0} \, F_A^2 (t_{min}) \;,
\eeq
where $T_A (b) = \int^{+\infty}_{-\infty} \mbox{d}z \, \rho_A ({\bg
  b}, z)$ is the nuclear normalized density profile, $\int
\mbox{d}^2b \,T_A (b) = 1$. The form factor $F_A$ is given by 
\beq
\label{eq:FA}
F_A (t_{min}) \;=\; \int \mbox{d}^2b \, J_0 (\sqrt{-t_{min}}b) \, T_A
(b) \;,
\eeq
where $t_{min} = -m_N^2 x^2_\Pom$, and $J_0(x)$ denotes the Bessel
function of the first kind. Strictly speaking, Eq.~(\ref{eq:sec}) is
valid for 
nuclear densities which depend separately on ${\bf b}$ and $z$,
however we have checked that calculations with an exact expression
lead to negligible corrections. Note that since Eq.~(\ref{eq:sec}) is
obtained under very general assumption, i.~e. analyticity and
unitarity, it can be applied for arbitrary values of $Q^2$ provided
$x$ is very small.
We have assumed $R_A^2 \gg R_N^2$, so
that the $t$-dependence of the $\vphot N$ cross section has been
neglected. For a deuteron, the double rescattering contribution has
the following form
\beq
\label{eq:deuteron}
\sigma_{\vphot d}^{(2)} \;=\; -2 \int_{-\infty}^{t_{min}} \mbox{d}t \,
\int^{M^2_{max}}_{M^2_{min}} \mbox{d}M^2 \,
\frac{\mbox{d}\sigma^\D_{\vphot N}}{\mbox{d}M^2 \mbox{d}t} \, F_D (t)
\;,
\eeq
where $F_D (t) = \exp (at)$, with $a = 40 \mbox{ GeV}^{-2}$
\cite{Cap98}. 

In Eqs.~(\ref{eq:sec}) and (\ref{eq:deuteron}), $M^2_{min}$ corresponds to
the minimal mass of the diffractively 
produced hadronic system, $M^2_{min} = 4m_{\pi}^2 = 0.08 \mbox{
  GeV}^2$, and $M^2_{max}$ is chosen according to the
condition: $x_\Pom \le x_\Pom^{max}$.
The choice of $x_\Pom^{max}$ is governed by the fact that the model is
only valid for $x_\Pom \ll 1$, i.~e. a large rapidity gap is required
in the experimental data. We use the standard choice for $x_\Pom^{max} = 0.1$
\cite{kaidalovrep}. It is convenient as it guarantees the disappearance of
nuclear shadowing at $x \sim 0.1$ as in experimental data.
Coherence effects are taken into account through $F_A (t_{min})$ in
Eq.~(\ref{eq:FA}), which is equal to 1 at $x \rightarrow 0$ and decreases with
increasing $x$ due to the loss of coherence for $x > x_{crit} \sim
\left( m_N R_A \right)^{-1}$. In the numerical calculations, a
3-parameter Woods-Saxon nuclear density profile with parameters from
\cite{jager} has been used. 

Higher order rescatterings in Eq.~(\ref{eq:sum}) are model dependent. We
use the Schwimmer unitarization \cite{schwimmer} for the total
$\vphot A$ cross section which is obtained from a summation of
fan-diagrams with triple-Pomeron interactions. It was checked in
\cite{Cap98} 
that it gives results very close to other reasonable models, such as
the quasi-eikonal model. The total cross section is then
\beq
\label{eq:sch}
\sigma_{\vphot A}^{Sch} \;=\; \sigma_{\vphot N} \, \int
\mbox{d}^2b \; \frac{A \, T_A (b)}{1 \,+\, (A-1) f(x, Q^2) T_A (b) } \;,
\eeq
where
\beq
\label{eq:f}
f(x,Q^2) \;=\; \frac{4\pi}{\sigma_{\vphot N}} \,
\int_{M^2_{min}}^{M^2_{max}} \mbox{d}M^2 \, \left[\frac{\mbox{d}
    \sigma^{\D}_{\vphot N}}{\mbox{d} M^2 \,
    \mbox{d}t} \right]_{t=0} \, F_A^2 (t_{min}) \;.
\eeq
Equation~(\ref{eq:sch})
does not take into account the shadowing effects of valence quarks nor
anti-shadowing effects, which may play an important role for $x \gsim
0.1$ \cite{Arn94}.

Nuclear shadowing is studied in terms of the ratios of
cross sections per nucleon for different nuclei, defined as
\beq
\label{eq:ratio}
R (A/B) \;=\; \frac{B}{A} \, \frac{\sigma_{\vphot A}}{\sigma_{\vphot B}} \;,
\eeq
as a function of $x$, which can in turn be expressed via structure
functions of the different nuclei. The simplest case is B$=$N, then
\beq
\label{eq:ratio2}
R^{Sch} \left( A/N \right) (x) \;=\; \int \mbox{d}^2b \, \frac{T_A
  (b)}{1 \,+\, (A-1) f(x, Q^2) T_A (b) } \;.
\eeq
In this framework shadowing can also be calculated at a fixed value of
the impact parameter $b$ for given values of $\left\{ x, Q^2 \right\}$
\beq
\label{eq:ratio3}
R^{Sch} \left( A/N \right) (b) \;=\; \frac{1}{1 \,+\, (A-1) f(x, Q^2)
  T_A (b) } \;.
\eeq
Thus the total $\vphot A$ cross section can be calculated within the
Glauber-Gribov model in a
parameter-free way provided the
total $\vphot N$ cross section and the differential cross section
for diffractive production are known.

\section{Diffractive parton densities from HERA}\label{sec:param}
The cross sections of the inclusive and diffractive processes are
expressed through the nucleon structure functions, which are in turn
associated with distribution functions of partons in the nucleon and
in the Pomeron. In DIS the structure function of a nucleon,
$F_2(x,Q^2)$, is
related to the total cross section of $\vphot N$ interaction through
\be
\sigma_{\vphot N} = \frac{4\pi^2 \alpha_{em}}{Q^2} F_2(x, Q^2) \;,
\ee
valid at small $x$. Similar to the inclusive DIS case, a factorization
theorem 
has been proven in perturbative QCD to hold for diffractive structure
functions \cite{collins}. The
relation between 
the diffractive cross section and the diffractive structure function
is given by
\be
\left[ \frac{\mbox{d}
    \sigma^{\D}_{\vphot N} (Q^2, x_\Pom, \beta)}{\mbox{d} M^2 \,
    \mbox{d}t} \right]_{t=0} \;=\; \frac{4\pi^2 \alpha_{em} \, B}{Q^2\,
  (Q^2 + M^2)} \, x_\Pom \, F_{2\D}^{(3)} (Q^2, x_\Pom, \beta) \,
\ee
where the usual factorization has been assumed:
\be
\frac{\mbox{d}
  \sigma^{\D}_{\vphot N} (x, Q^2, M^2, t)}{\mbox{d} M^2 \,
  \mbox{d}t} \;=\; \left[\frac{\mbox{d}
    \sigma^{\D}_{\vphot N} (x, Q^2, M^2)}{\mbox{d} M^2 \,
    \mbox{d}t} \right]_{t=0} \, e^{Bt} \;.
\ee
A further assumption about the lower part of Fig.~\ref{fig:diag2}, the
so-called Regge factorization \cite{adloff97}, allows us to write the diffractive
structure function as
\beq
\label{eq:reggefact}
F_{2\D}^{(3)}(x_\Pom, Q^2, \beta) \;=\; f_\Pom(x_\Pom) \, F(\beta,Q^2) \;. 
\eeq
The first factor is referred to as the ($t$-integrated) Pomeron flux
(we will not take into account the sub-leading contributions from
other reggeon trajectories), which is in turn defined as
\be
f_\Pom(x_\Pom) \;=\; A_\Pom \int^{t_{min}}_{t_{cut}} \, \frac{e^{B_0 \,
    t}}{x_\Pom^{2\alpha_P (t) \,-\, 1}} \, \mbox{d}t \;, 
\ee
where $A_\Pom$ is a constant that fixes the normalization of the flux,
and $t_{cut} = -1 \, \mbox{GeV}^2$.
As usual, we assume a linear Pomeron trajectory, $\alpha_\Pom \left( t
\right) =
\alpha_\Pom (0) + \alpha_\Pom' t$. The second factor in Eq.~(\ref{eq:reggefact}),
$F(\beta,Q^2)$, is the Pomeron structure function. For the sake of
simplicity, Eq.~(\ref{eq:f}) can now be reduced to
\beq
\label{eq:fsimpl}
f(x, Q^2) \;&=&\; 4\pi\; \int_x^{x_\Pom^{max}} \mbox{d}x_\Pom \,B(x_\Pom)\,
\frac{F_{2 \D}^{(3)}(x_\Pom, Q^2, \beta)}{F_2 (x, Q^2)} \; F_A^2
(t_{min.}) \;,
\eeq
where $B(x_\Pom) = B_0 \,+\, \alpha_\Pom' \ln \frac{1}{x_\Pom}$. The
ratio of structure functions in the integrand 
can be understood as the density of partons in the
Pomeron compared to the density of partons in the nucleon.

The determination of the
distribution of partons in the nucleon and Pomeron cannot be carried
out in pQCD, since it 
depends on soft processes such as confinement. The inclusive distributions are
measured in DIS experiments to high accuracy and predictions of
the DGLAP evolution are in good agreement with experiment for a broad
range of $x$ and $Q^2$ values. Yet, the experimental status of the
diffractive distribution functions was, until recently, uncertain. The
biggest uncertainty was related to the gluon distribution, because it
is not measured directly in the experiment.
Some authors therefore assumed that the 
shadowing from quarks and gluons are of equal strength
\cite{cap99,gyulassy91,eskola98}, while others 
predicted that the gluon shadowing dominates
\cite{huang98,hammon98,fgs03}.

The results of new high-precision measurements  of
the diffractive parton distribution functions (DPDFs) presented 
by the H1 Collaboration \cite{H1new1,H1new2},
shed new light on the role of gluon shadowing at intermediate
$Q^2$. The quark singlet and gluon DPDFs were fitted by a simple
function
\beq
\beta \, f^\D_i(\beta, Q_0^2) \;&=&\;  A_i \beta^{-B_i} \left(1-\beta
\right)^{C_i} \\
\beta \, g^\D (\beta,Q_0^2) \;&=&\; A_g \beta^{-B_g} \left(1-\beta
\right)^{C_g} \;,
\label{eq:fitfunc}
\eeq
where $A_i$, $B_i$ and $C_i$ are fitting parameters, at a given
$Q_0^2 \sim 2 \mbox{ GeV}^2$. In general, to
ensure that the r.h.s. of Eq.~(\ref{eq:fitfunc}) always disappears as
$\beta \rightarrow 1$, it 
should be multiplied by $\exp \left( \frac{-0.01}{1-x}
\right)$. QCD evolution to arbitrary $Q^2$ was performed by the H1 group.
For the quark singlet distribution, all three parameters $\{
A_q,B_q,C_q\}$ were used
in the fit, while the gluon density is found to be insensitive to the
$B_g$ parameter, which is therefore set to zero. This fit is referred
to as the 'FIT A'. Additionally,
because of
DGLAP evolution, data at $\beta \geq 0.3$ cannot
constrain the gluon density because of the large quark contribution to
the $Q^2$ evolution. This lack of sensitivity is confirmed by
repeating the fit with the parameter $C_g$ set to zero, and is further
referred to as the 'FIT B'. The values of the corresponding Pomeron
parameters are listed in Tab.~\ref{tab:parameters}. The fitted quark
singlet distributions of 'FIT A'
and 'FIT B' are consistent with each other.

Inclusion of diffractive
di-jet production in the analysis provides a big improvement in the
precision of the gluon density measurement \cite{h1mozer},which can be further
constrained by the diffractive production of charm in
DIS \cite{h1kapichine}. A combined fit to DDIS data and 
diffractive di-jets \cite{h1mozer} results in a curve similar
to fit B (described above) yet with a slightly smaller gluon density
at $\beta > 0.5$. For completeness, we also
compare the new results with the old H1 parameterization \cite{H1abs}
presented in 2002 (corresponding Pomeron parameters are given in
Table \ref{tab:parameters}).  

The gluon distribution is
almost a factor of 10 bigger than the quark distribution at the same $Q^2$
for a broad region of $\beta$ \cite{H1new1,H1new2}. Shadowing from quarks was
calculated in \cite{armesto03} using the old H1 parameterizations, and
we have checked that the new H1 data are consistent with their
calculations. Furthermore, in the relevant kinematical range for
hadron-nucleus and nucleus-nucleus collisions at RHIC and
LHC, the gluon density dominates. In what follows, we 
will therefore consider structure functions of gluons in nuclei. The
first term in Eq.~(\ref{eq:sum}) is then proportional to $A G_N(x,Q^2)$,
while the second rescattering term is correspondingly equal to
$-A(A-1)G_N(x,Q^2) \, \int\mbox{d}^2T_A(b)^2 \,f(x,Q^2)$, where
\beq
\label{eq:fgluon}
f \left(x, Q^2 \right) \;=\; 4\pi \, \int^{x_\Pom^{max}}_x
\mbox{d}x_\Pom \, B(x_\Pom) f_\Pom(x_\Pom) \, \frac{ \beta
  g^\D (\beta,Q^2)}{G_N (x,Q^2)} \, F_A^2 (t_{min}) \;.
\eeq
Thus the gluon shadowing factor, $R_g$, is calculated according to
Eq.~(\ref{eq:ratio2}) with $f(x,Q^2)$ given by
Eq.~(\ref{eq:fgluon}). The gluon distribution of the nucleon,
$G_N(x,Q^2) = xg(x,Q^2)$, was taken from CTEQ6M parameterization \cite{cteq6}.
\begin{table}[t!]
  \begin{center}
    \begin{tabular}{cc}
      \hline
      \hline
      $\alpha_\Pom (0)$ [FIT A]     & 1.118 \\
      $\alpha_\Pom (0)$ [FIT B]     & 1.111 \\
      $\alpha_\Pom'$                & 0.06 $\mbox{GeV}^{-2}$ \\
      $B_0$                         & 5.5 $\mbox{GeV}^{-2}$ \\
      \hline
      $\alpha_\Pom (0)$ [2002] & \it{1.173} \\
      $\alpha_\Pom'$ [2002]    & \it{0.26 $\mbox{GeV}^{-2}$} \\
      $B_0$ [2002]             & \it{4.6 $\mbox{GeV}^{-2}$} \\
      \hline
      \hline
    \end{tabular}
    \caption{Pomeron parameters from H1 Collaboration \cite{H1new1,H1new2,H1abs}.}
    \label{tab:parameters}
  \end{center}
\end{table}

The model for diffractive production by virtual photon described
above
is applicable for intermediate $Q^2 > 1-2
\mbox{ GeV}^2$ \cite{H1new1,H1new2}. It is also important to keep in mind that the
parameterization of H1 data 
leads to a violation of unitarity for $x \rightarrow 0$ and should be
modified at very low x (probably already at $x \sim 10^{-4}$ for low
$Q^2$) \cite{kkmr2003}. Therefore our predictions are reliable in the
region $x > 10^{-4}$ which is relevant for RHIC and most experiments
at LHC, and should be taken with care for $x<10^{-4}$. 

\section{Numerical results}\label{sec:results}
\begin{figure}
  \centering
  \includegraphics[scale=.8]{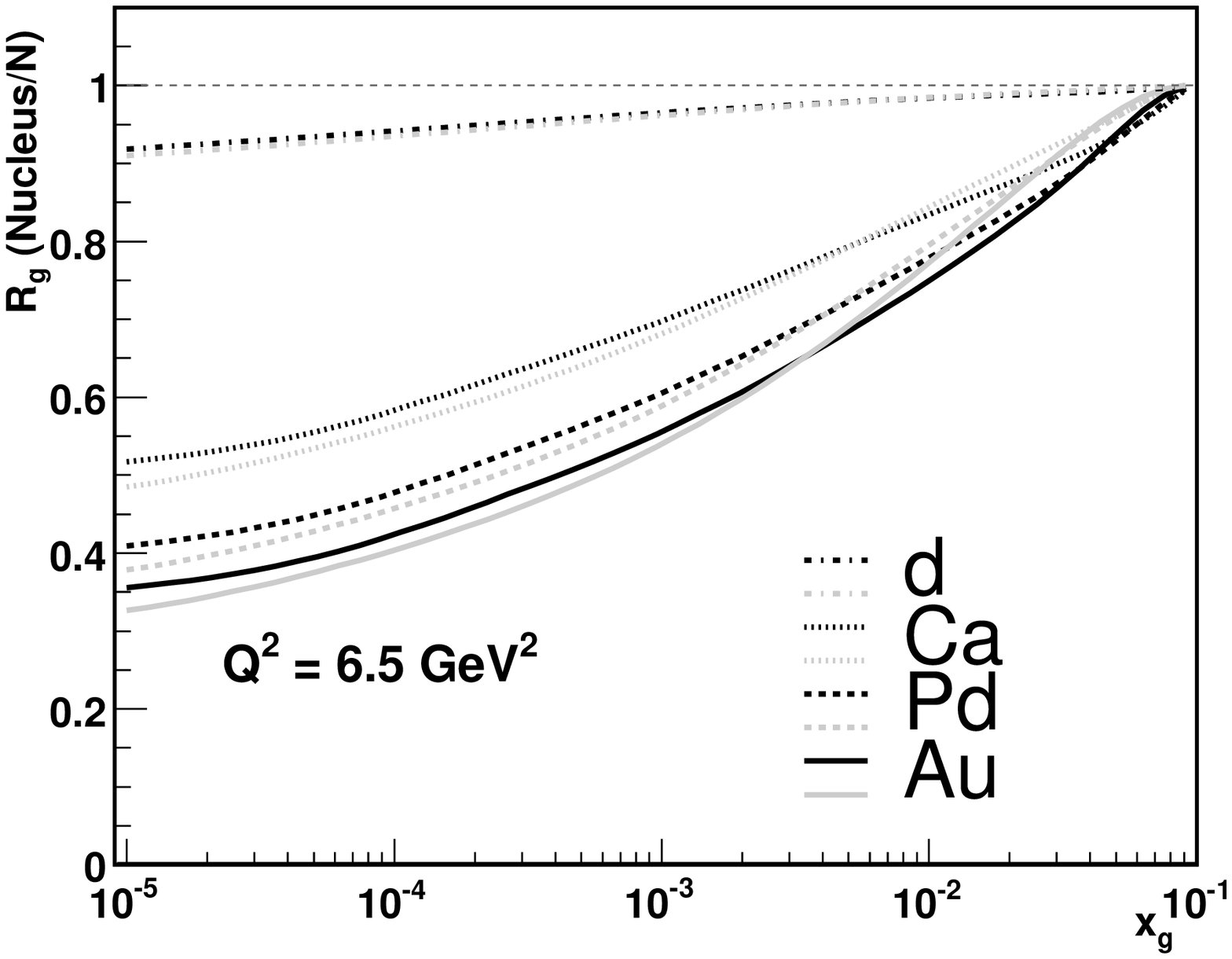}
  \caption{Gluon shadowing for deuteron (dash-dotted curves) and for
    heavy ions: Ca (dotted curves), Pd (dashed 
    curves), Au (solid curves). Black curves are for FIT A and grey curves
    are for FIT B.}
  \label{fig:AN}
\end{figure}
\begin{figure}[ht]
  \begin{center}
    \includegraphics[scale=.8]{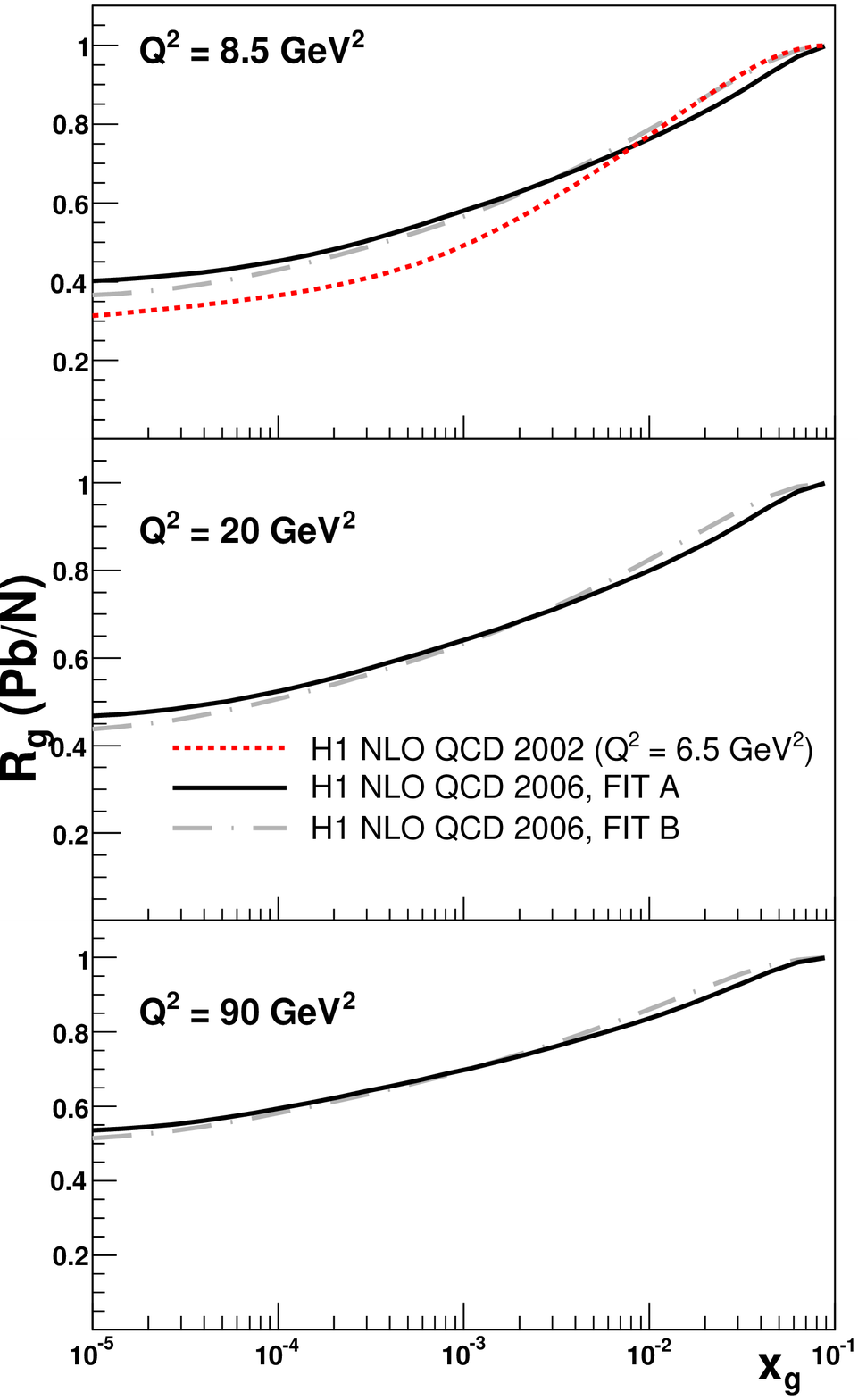}
    \caption{Gluon shadowing for Pb for different virtualities, $Q^2$.}
    \label{fig:PbQ2}
  \end{center}
\end{figure}
Gluon shadowing for deuteron and various heavy ions (Ca, Pd and Au) at
$Q^2 = 6.5$ 
GeV$^2$ is presented in
Fig.~\ref{fig:AN}. We have made the calculations using two fits of the
gluon diffractive distribution function, FIT A and FIT B.
The gluon shadowing is very strong at small $x$, and disappearing at
$x = x_\Pom^{max}$. This is a
consequence of the coherence effect in the form factor Eq.~(\ref{eq:FA}),
and the vanishing integration domain in Eq.~(\ref{eq:fsimpl}). Gluon
shadowing is as low as 0.35 for the Au/nucleon ratio. In
Fig.~\ref{fig:PbQ2} shadowing in Pb is shown for different values of
$Q^2$. We have also included the result using the 2002 H1
parameterization for $Q^2 = 6.5$ GeV$^2$. Although the difference for
the experimentally measured gluon distribution function is quite
substantial, the shadowing
contribution is stable, changing maximally 30$\%$ for the lowest
$x$.

There is little change to be observed in the gluon shadowing ratio
with increasing $Q^2$. The QCD evolution of the main term and
rescattering terms are effectively treated separately in our approach,
and therefore the shadowing correction has a slow, logarithmic
dependence on $Q^2$ \cite{Bro02}. The $Q^2$ dependence of shadowing was studied
within a similar model in \cite{Cap98}
and good agreement with existing data was found.

\begin{figure}[ht]
  \begin{center}
    \includegraphics[scale=.8]{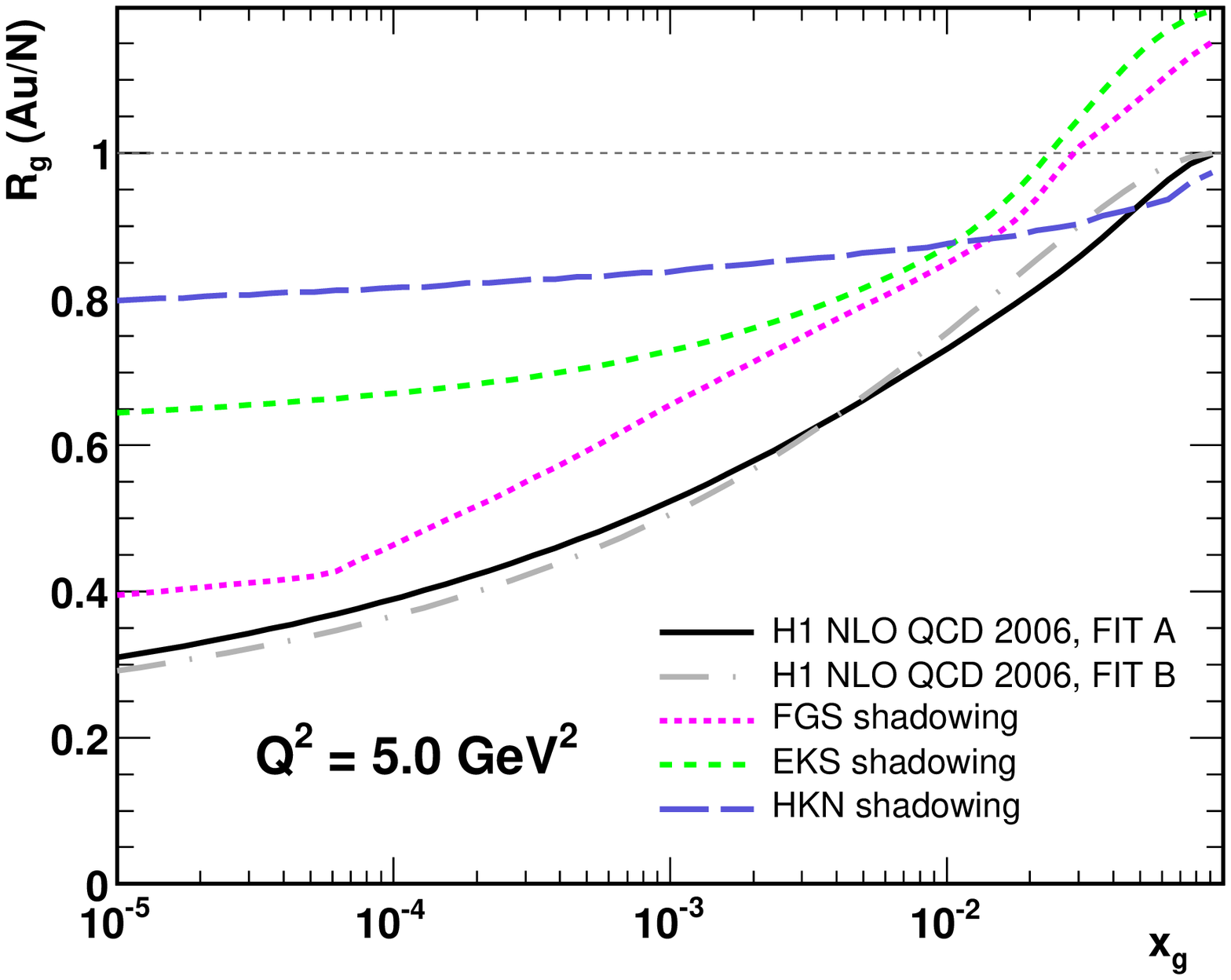}
    \caption{Comparison of the results of the Glauber-Gribov model
      with FGS model \cite{fgs03}, EKS
      \cite{eks98} and HKM \cite{hkm04} parameterizations.}
    \label{fig:comp}
  \end{center}
\end{figure}
A comparison of the results of our model with other predictions is
presented in Fig.~\ref{fig:comp} for Pb at $Q^2 = 5$ GeV$^2$. For all $x
\leq 10^{-1}$ our model predicts slightly stronger gluon shadowing compared to
FGS 
\cite{fgs03}, while models based on global fits to existing data on nuclear
modifications and DGLAP evolution \cite{eks98,hkm04} predict a modest
gluon shadowing effect. 
The authors of
\cite{fgs03} have made calculations in a similar framework as the
presented model, using a quasi-eikonal summation of higher-order
diagrams. In general, the quasi-eikonal model gives rise to a stronger
shadowing effect than models including enhanced diagrams, i.e. Pomeron
interactions, like the Schwimmer model, at $x<10^{-3}$. This is not so
clear from the figure as the authors of \cite{fgs03} put $x_\Pom^{max}
= 0.03$ and include anti-shadowing for gluons. The quasi-eikonal model
will, at asymptotical energies, lead to the grey disc limit.

\begin{figure}[ht]
  \begin{center}
    \includegraphics[scale=.8]{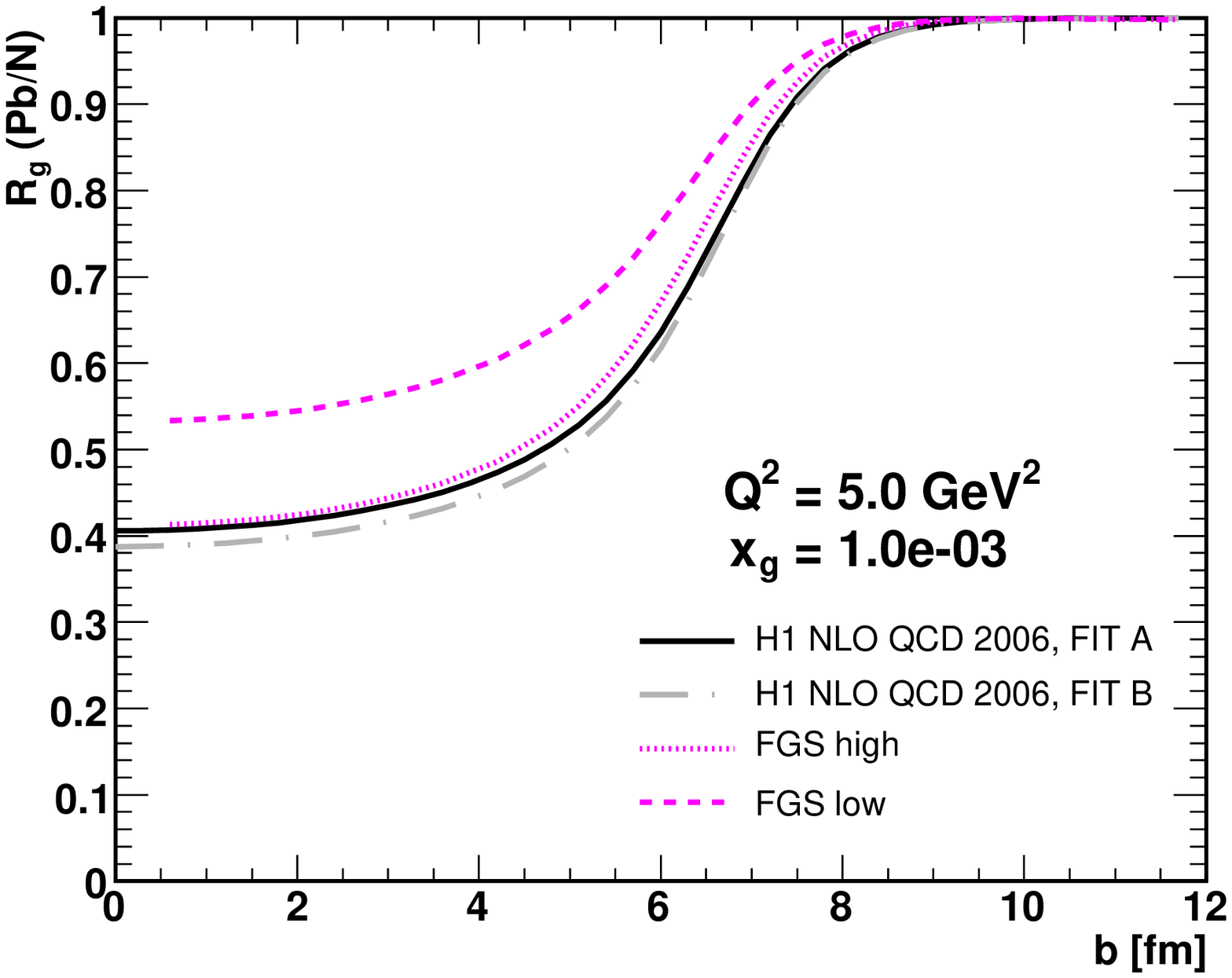}
    \caption{Impact parameter dependence of gluon shadowing in the
      Glauber-Gribov model, compared to the predictions of FGS
      \cite{fgs03}.}
    \label{fig:Imp}
  \end{center}
\end{figure}
We have also calculated the impact parameter dependence of
Glauber-Gribov gluon shadowing given $x = 10^{-3}$ and $Q^2 = 5$
GeV$^2$, and compared it to results from \cite{fgs03} in both ``high''
and ``low'' gluon shadowing mode. Our model predict
shadowing in agreement with the strongest FGS prediction, which is
20\% lower than in the ``low'' gluon shadowing mode for central impact
parameters. The 
shape seems to be compatible between the two models. 
We have also
checked that agreement with the FGS model improves when updated H1
parameterizations are used \cite{vad07}.

\section{Conclusions}\label{sec:concl}
Calculations of gluon shadowing in heavy-ions and deuteron have been
made within the Gribov-Glauber model, which is a parameter-free
framework to calculate shadowing effects in hadron-nucleus collisions
at high energy. Diffractive structure functions
have been parameterized using the latest NLO QCD fits from HERA
experiments. Rather substantial shadowing effects are predicted for
heavy nuclei at $x<10^{-3}$, which is slowly changing with increasing
$Q^2$. This will have implications for both light and heavy particle
production in hadron-nucleus 
collisions at RHIC and LHC, where the relevant $\left\{ x, Q^2 \right\}$ range
is being probed. 

Comparison to other models predicting strong shadowing effects at $x <
10^{-2}$ have been performed.
Calculations of Glauber-Gribov shadowing using Schwimmer summation of
fan diagrams differ from other approaches within the same
framework mainly due to differences in the treatment of higher order
rescatterings, which is important in the low-$x$ region,
and due to differences in the QCD evolution in $\log Q^2$. The
advantages of our method is the clear treatment of these effects
within a well-established framework.

The nuclear PDFs can be measured in ultra-peripheral
collisions at both RHIC and LHC \cite{strik07}. In the future,
hopefully an electron-ion 
collider (eRHIC) would provide clear information about DIS on nuclei
which would dramatically improve our understanding of high-energy
nuclear effects such as shadowing.

\section*{Acknowledgments}

The authors would like to thank N.~P.~Armesto,
A.~Capella, V.~Guzey and M.~Strikman for interesting discussions, and V.~Guzey for
providing updated curves for the
figures. This work was supported by the Norwegian 
Research Council (NFR) under contract No. 166727/V30, QUOTA-program,
RFBF-06-02-17912, RFBF-06-02-72041-MNTI, INTAS 05-103-7515, grant of
leading scientific schools 845.2006.2 and support of Federal 
Agency on Atomic Energy of Russia.

\vfill\eject
\end{document}